\documentclass[twocolumn,aps,prl,superscriptaddress,showpacs,floatfix]{revtex4}
\usepackage{graphicx}

\begin{document}

\title{Momentum dependence of the symmetry potential and 
nuclear reactions induced by neutron-rich nuclei at RIA}
\author{Bao-An Li}
\affiliation{Department of Chemistry and Physics, P.O. Box 419, Arkansas State
University, State University, Arkansas 72467-0419, USA}
\author{Champak B. Das}
\affiliation{Physics Department, McGill University, Montr{\'e}al, Canada H3A 2T8}
\author{Subal Das Gupta}
\affiliation{Physics Department, McGill University, Montr{\'e}al, Canada H3A 2T8}
\author{Charles Gale}
\affiliation{Physics Department, McGill University, Montr{\'e}al, Canada H3A 2T8}
\date{\today}

\begin{abstract}
Effects of the momentum-dependence of the symmetry potential in nuclear reactions
induced by neutron-rich nuclei at RIA energies are studied using an 
isospin- and momentum-dependent transport model. It is found that symmetry 
potentials with and without the momentum-dependence but corresponding to the same
density-dependent symmetry energy $E_{sym}(\rho)$ lead to significantly different predictions 
on several $E_{sym}(\rho)$-sensitive experimental observables. The momentum-dependence of the
symmetry potential is thus critically important for investigating accurately the 
equation of state (${\rm EOS}$) and novel properties of dense neutron-rich matter at RIA.
\end{abstract}

\maketitle

The rapid advance in technologies to accelerate radioactive beams has opened up 
several new frontiers in nuclear sciences\cite{nsac02,tani01}. 
Particularly, the high energy radioactive beams to be available 
at the planned Rare Isotope Accelerator (RIA) and the new 
accelerator facility at GSI provide a great opportunity to explore the 
${\rm EOS}$ and novel properties of dense neutron-rich matter\cite{ireview98,ibook01,dan02,li03}. 
The energy per particle $E(\rho ,\delta )$ in asymmetric nuclear 
matter of density $\rho$ and relative neutron excess 
$\delta =(\rho_{n}-\rho _{p})/(\rho _{p}+\rho _{n})$ is usually expressed as
$E(\rho ,\delta )=E(\rho ,\delta =0)+E_{\text{\textrm{sym}}}(\rho )\delta
^{2}+{\cal O}(\delta^4)$,  
where $E_{\rm sym}(\rho)$ is the density-dependent nuclear symmetry energy. 
The latter is among the most important and yet very poorly known properties of dense 
neutron-rich matter\cite{bethe,lat01}. For instance, it is 
important for Type II supernova explosions, 
for neutron-star mergers, and for the stability of neutron stars. It also 
determines the proton fraction and electron chemical potential in neutron stars at 
$\beta$ equilibrium. These quantities consequently determine the cooling 
rate and neutrino emission flux of protoneutron stars and the possibility
of kaon condensation in dense stellar matter\cite{bethe,lat01}. 
In nuclear reactions induced by neutron-rich nuclei, the $E_{sym}(\rho)$ reveals itself
through dynamical effects of the corresponding symmetry potentials acting differently 
on neutrons and protons. 
Based on isospin-dependent transport model calculations, several experimental 
observables have been identified as promising probes of the $E_{sym}(\rho)$, 
such as, the neutron/proton ratio \cite{li97,tan01,bar02}, the neutron-proton 
differential flow \cite{li00,gre03}, the neutron-proton correlation function\cite{chen03}
and the isobaric yield ratios of light clusters\cite{chen03b}. 
However, in all existing transport models the momentum-dependence of the 
symmetry (isovector) potential has never been
taken into account. Effects of the momentum-dependence of the symmetry  
potential in nuclear reactions are thus completely unknown, although effects of 
the momentum-dependence of the isoscalar potential are 
well-known\cite{gbd87,gale90,pan93,zhang94}. 
This is mainly because only very recently was the momentum-dependence of the symmetry potential 
given in a form practically usable in transport model calculations\cite{das03}. 
In this work, we study for the very first time effects of the momentum-dependence 
of the symmetry potential within an isospin- and momentum-dependent transport model 
for nuclear reactions induced by neutron-rich nuclei at RIA energies. It is found 
that symmetry potentials with and without 
the momentum-dependence but corresponding to the same symmetry energy $E_{sym}(\rho)$ 
lead to significantly different predictions on several sensitive probes of the $E_{sym}(\rho)$. 
Moreover, these observables are more sensitive to the variation of $E_{sym}(\rho)$ 
in calculations with the momentum-dependent symmetry potentials. 

That the momentum-dependence of the single particle potential is different for neutrons and protons 
in asymmetric nuclear matter is well-known and has been a subject of intensive research based on various 
many-body theories using non-local interactions, see e.g., ref. \cite{bom1} for a review. 
Guided by a Hartree-Fock calculation using the Gogny effective interaction, the single nucleon 
potential was recently parameterized as \cite{das03} 
\begin{eqnarray}\label{mdi}
U(\rho,\delta,\vec p,\tau) &=& A_u\frac{\rho_{\tau'}}{\rho_0}
+A_l\frac{\rho_{\tau}}{\rho_0}\nonumber \\ &+&
B(\frac{\rho}{\rho_0})^{\sigma}(1-x\delta^2)-8\tau x\frac{B}{\sigma+1}\frac{\rho^{
\sigma-1}}{\rho_0^{\sigma}}\delta\rho_{\tau'} \nonumber \\
&+&\frac{2C_{\tau,\tau}}{\rho_0}
\int d^3p'\frac{f_{\tau}(\vec r,\vec p')}{1+(\vec p-\vec p')^2/\Lambda^2}\nonumber \\
&+&\frac{2C_{\tau,\tau'}}{\rho_0}
\int d^3p'\frac{f_{\tau'}(\vec r,\vec p')}{1+(\vec p-\vec p')^2/\Lambda^2}
\end{eqnarray}
In the above $\tau=1/2$ ($-1/2$) for neutrons (protons) and $\tau\neq\tau'$; $\sigma=4/3$;
$f_{\tau}(\vec r,\vec p)$ is the phase space distribution 
function at coordinate $\vec{r}$ and momentum $\vec{p}$. 
The parameters $A_u, A_l, B, C_{\tau,\tau}, C_{\tau,\tau'}$ and $\Lambda$ listed 
in ref.\cite{das03} were obtained by fitting the momentum-dependence of the 
$U(\rho,\delta,\vec p,\tau)$ predicted by the Gogny Hartree-Fock and/or the Brueckner-Hartree-Fock 
calculations, saturation properties of symmetric nuclear matter and the symmetry energy of 30 MeV 
at normal nuclear matter density $\rho_0=0.16/fm^3$. The compressibility of symmetric 
nuclear matter $K_{0}$ is set to be 211 MeV. The momentum-dependence of the symmetry potential 
steams from the different interaction strength parameters $C_{\tau,\tau'}$ and $C_{\tau,\tau}$ 
for a nucleon of isospin $\tau$ interacting, respectively, with unlike and like nucleons in 
the background fields. More specifically, $C_{unlike}=-103.4$ MeV while $C_{like}=-11.7$ MeV. 

The parameter $x$ was introduced to reflect the largely uncertain behavior of $E_{sym}(\rho)$, 
specifically its potential part $E_{sym}^p(\rho)$.  
The latter corresponding to $x=1$ (denoted by MDI(1) which gives the same $E_{sym}(\rho)$
as the default Gogny interaction) and 
$x=0$ (MDI(0)) together with the kinetic contribution 
$E_{sym}^k(\rho)=\hbar^2/6m_n(3\pi^2\rho/2)^{2/3}$ are shown in Fig.\ 1. 
\begin{figure}[ht]
\includegraphics[scale=0.5,angle=-90]{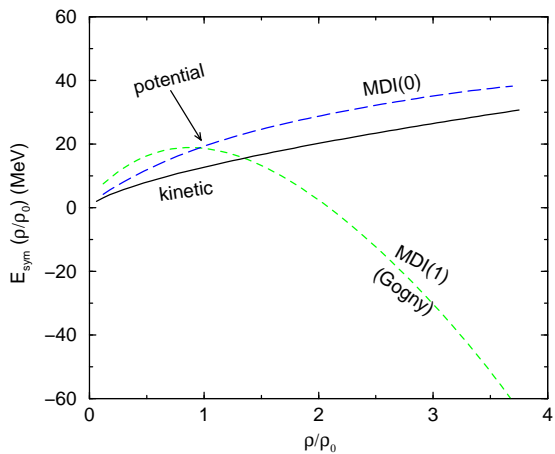}
\caption{{\protect\small The density dependence of the 
potential and kinetic parts of nuclear symmetry energy.}}
\label{esym}
\end{figure}
The variation of $E_{sym}(\rho)$ resulted by changing from the MDI(1) to the MDI(0) parameter set
is well within the uncertain range predicted by various many-body theories, 
see e.g., refs.\cite{brown,vmb,bom91}.   
The potential contribution to the symmetry energy can be parameterized as 
\begin{equation}\label{mdi1}
E_{sym}^p(\rho)=3.08+39.6u-29.2u^2+5.68u^3-0.523u^4
\end{equation}   
for the MDI(1) (Gogny) and
\begin{equation}\label{mdi0}
E_{sym}^p(\rho)=1.27+25.4u-9.31u^2+2.17u^3-0.21u^4
\end{equation}   
for the MDI(0), where $u\equiv \rho/\rho_0$ is the reduced nucleon density. 

We implemented the single particle potential in eq. \ref{mdi} in an isospin-dependent transport
model\cite{li97}. Since the C-terms in the single particle potential depend inseparably on the density,
momentum and isospin, to investigate effects of the momentum-dependence of the symmetry potential we shall
compare results obtained using eq. \ref{mdi} with those obtained using a single nucleon potential 
$U_{noms}(\rho,\delta,\vec{p},\tau)\equiv U_0(\rho,\vec{p})+U_{sym}(\rho,\delta,\tau)$ that
has almost the same momentum-dependent isoscalar part $U_0(\rho,\vec{p})$ as that embedded 
in eq. \ref{mdi} and a momentum-independent symmetry potential $U_{sym}(\rho,\delta,\tau)$ 
that gives the same $E_{sym}(\rho)$ as eq.\ \ref{mdi}.
The $U_{sym}(\rho,\delta,\tau)$ can be obtained 
from $U_{sym}(\rho,\delta,\tau)=\partial W_{sym}/\partial \rho_{\tau}$, where $W_{sym}$ is
the isospin-dependent part of the potential energy density 
$W_{sym}=E_{sym}^p(\rho)\cdot\rho\cdot\delta^2$. Using the 
parameterizations for $E_{sym}^p(\rho)$ in eqs. \ref{mdi1} and \ref{mdi0}, we obtain
\begin{eqnarray}
&&U^{MDI1}_{sym}(\rho,\delta,\tau)=4\tau\delta(3.08+39.6u-29.2u^2 +5.68u^3
\nonumber \\ && -0.52u^4)
-\delta^2(3.08+29.2u^2-11.4u^3+1.57u^4)
\end{eqnarray} 
for the MDI(1) parameter set and 
\begin{eqnarray}
&&U^{MDI0}_{sym}(\rho,\delta,\tau)=4\tau\delta(1.27+25.4u-9.31u^2 +2.17u^3
\nonumber \\ && -0.21u^4)
-\delta^2(1.27+9.31u^2-4.33u^3+0.63u^4)
\end{eqnarray}  
for the MDI(0) parameter set.
To identify reliably effects of the momentum-dependence of the symmetry potential  
without much interference from density effects the $U_{noms}(\rho,\delta,\vec{p},\tau)$ 
should lead to almost the same reaction dynamics 
and evolution of density profiles as the single particle potential in eq.\ \ref{mdi}.
Both of them are mainly determined by the isoscalar potential for which 
we select the original MDYI interaction\cite{gale90} 
\begin{eqnarray}\label{mdyi}
U_0(\rho,\vec{p})&=&-110.44u+140.9u^{1.24}\nonumber \\
&&-\frac{130}{\rho_0}
\int d^3p'\frac{f(\vec r,\vec p')}{1+(\vec p-\vec p')^2/(1.58p_F^0)^2},
\end{eqnarray}
where $p_F^0$ is the Fermi momentum. The compressibility $K_0$ for this interaction is 215 MeV.
We compared numerically $U_0(\rho,\vec{p})$ in eq. \ref{mdyi} with 
$(U_n(\rho,\delta,\vec p)+U_p(\rho,\delta,\vec p))/2$ obtained from eq. 1. They are indeed 
very close and both agree with the nucleon optical potential data.
We also verified numerically that the potential $U_{noms}(\rho,\delta,\vec{p},\tau)$ constructed this way 
leads to about the same reaction dynamics and the evolution of density profiles 
as the potential in eq. \ref{mdi}. 
\begin{figure}[ht]
\includegraphics[scale=0.5,angle=-90]{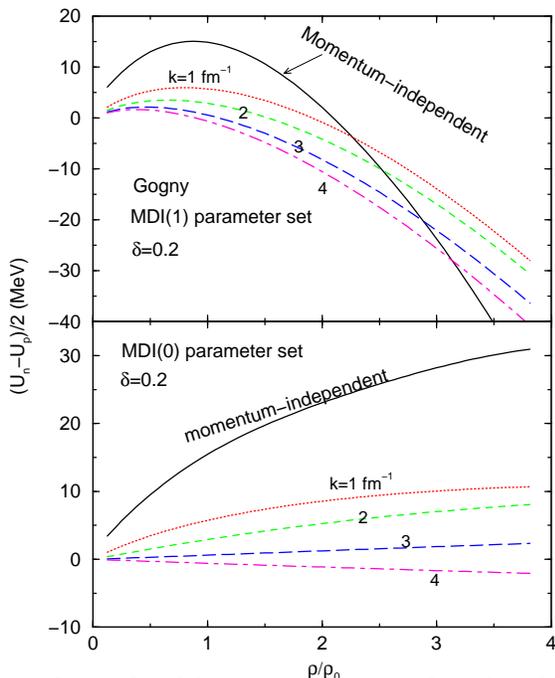}
\vspace{-0.5cm}
\caption{{\protect\small.Strengths of the symmetry potentials with and without (solid) 
the momentum-dependence as a function of density as measured by the difference 
between neutron and proton potentials.}}
\label{usym}
\end{figure} 

The strengths of the symmetry potentials with and without the momentum-dependence 
but corresponding to the same $E_{sym}(\rho)$ are compared in Fig.\ 2 
by examining the difference between neutron and proton potentials $(U_n-U_p)/2$. 
The symmetry potential without the momentum-dependence is 
higher in magnitude and has generally steeper slopes than the 
momentum-dependent one for $\rho/\rho_0\leq 2.3$ 
with the MDI(1) parameter set and at all densities for the MDI(0) parameter set.
Moreover, the strength of momentum-dependent symmetry potentials decreases with 
the increasing momentum. Thus the difference between
the symmetry potentials with and without the momentum-dependence is larger for nucleons
with higher momenta. Several experimental observables are known to be sensitive 
only to the symmetry potential but not to the isoscalar potential. 
These observables are mainly neutron-proton differential or relative 
quantities\cite{li97,li00} where effects of the isoscalar potential 
with or without the momentum-dependence 
are largely canceled out. These observables are also insensitive 
to the in-medium nucleon-nucleon (NN) cross sections\cite{ireview98,li97,chen03}.
We use here the isospin-dependent experimental NN cross sections.
Assuming the symmetry potential is 
momentum-independent, these observables were previously proposed 
as promising probes of the density-dependence of the symmetry energy. 
We compare in the following several such observables calculated with 
the $U(\rho,\delta,\vec p,\tau)$ in eq. \ref{mdi} and the 
$U_{noms}(\rho,\delta,\vec{p},\tau)$ corresponding to 
the MDI(1) parameter set. Conclusions based on calculations 
using the MDI(0) parameter set are the same\cite{li04}.
 
Shown in the upper window of Fig.\ 3 is the average isospin asymmetry 
$\delta_{free}(y)$ of free nucleons as a function of rapidity $y$ from 1400 events
of $^{132}Sn+^{124}Sn$ reactions at a beam energy of 400 MeV/nucleon and an 
impact parameter of 5 fm. We define here free nucleons as those with local 
baryon densities less than $\rho_0/8$. The value of $\delta_{free}(y)$ reflects 
mainly the degree of isospin fractionation between the free nucleons and the 
bound ones at freeze-out. It is also influenced slightly by the 
production of more $\pi^-$ than $\pi^+$ 
mesons in the reaction. The momentum-independent symmetry potential leads to
significantly higher value of $\delta_{free}(y)$ than the momentum-dependent one. Moreover, the
difference tends to increase with rapidity. At midrapidity the predicted $\delta_{free}(y)$ 
values are close to the value expected when a complete isospin 
equilibrium is established among all target and projectile nucleons.
These features are what one expects from the strength of the symmetry potentials as shown in 
the upper window of Fig.\ 2. The generally repulsive (attractive) symmetry potential 
for neutrons (protons) around $\rho_0$ causes more neutrons (protons) to be free (bound). 
The momentum-independent symmetry potential is higher and steeper than 
the momentum-dependent one and the difference between them increases with momentum, 
it thus leads to the higher $\delta_{free}(y)$ values especially at higher rapidities.

\begin{figure}[ht]
\includegraphics[scale=0.5,angle=-90]{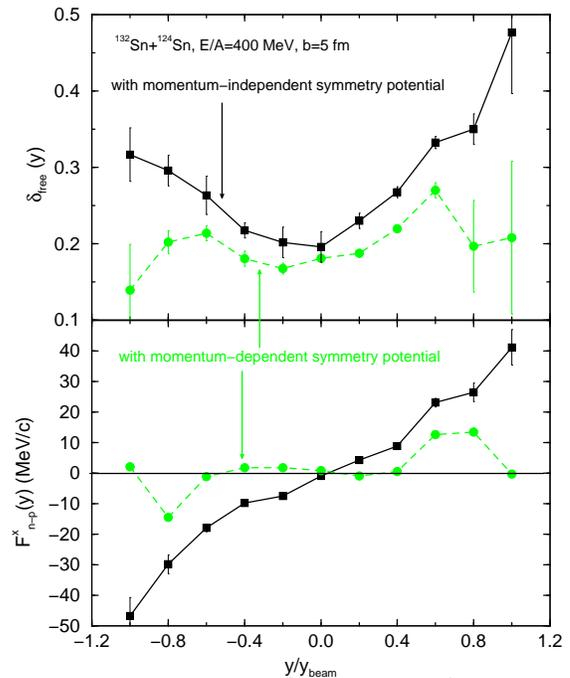}
\vspace{-0.5cm}
\caption{{\protect\small Isospin-asymmetry (upper window) and neutron-proton differential flow 
(lower window) of free nucleons as a function of rapidity. The solid (dashed) lines are calculated with the
momentum-independent (-dependent) symmetry potential.}}
\label{dflow}
\end{figure}
Shown in the lower window of Fig.\ 3 is the neutron-proton differential flow
$F^x_{n-p}(y)\equiv\sum_{i=1}^{N(y)}(p^x_iw_i)/N(y)$, where $w_i=1 (-1)$ for neutrons (protons) 
and $N(y)$ is the total number of free nucleons at rapidity $y$. The differential flow combines 
constructively effects of the symmetry potential on the isospin fractionation and the collective flow. 
It has the advantage of maximizing effects of the symmetry potential while minimizing effects
of the isoscalar potential\cite{li00}. Compared to the momentum-dependent symmetry potential embedded
in eq.\ \ref{mdi}, the momentum-independent symmetry potential 
$U^{MDI1}_{sym}(\rho,\delta,\tau)$ makes not only more neutrons free but also gives them 
higher transverse momenta in the reaction plane because of its higher magnitude and steeper 
density slopes. As a result, the differential flow
with the momentum-independent symmetry potential is significantly higher. 

Complementary information about how the symmetry potential depends on the momentum 
can be obtained by studying the ratio of neutrons to protons as a function of their 
transverse momentum $p_t$. Shown in Fig.\ 4 is this ratio around the midrapidity within
$|y_{cms}/y_{beam}|\leq 0.3$. The overall rise of the ratio at low $p_t$ is due to the 
Coulomb force which shifts protons from low to higher energies. 
It is seen that the difference between the predicted ratios increases with $p_t$,
reflecting the increasingly larger difference between the symmetry potentials with and
without the momentum-dependence for nucleons with higher momenta as shown in Fig.\ 2. 
The high $p_t$ nucleons are thus more useful for studying the momentum-dependence 
of the symmetry potential. 

\begin{figure}[ht]
\includegraphics[scale=0.45,angle=-90]{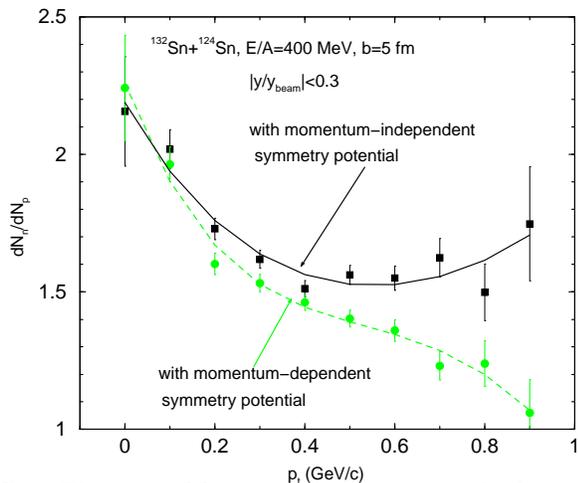}
\vspace{-.5cm}
\caption{{\protect\small The ratio of free neutron to proton multiplicity as a function of transverse
momentum at midrapidity. The solid (dashed) line is calculated with the 
momentum-independent (-dependent) symmetry potential.}}
\label{npratio}
\end{figure}
Are the observables studied above still sensitive to the variation of $E_{sym}(\rho)$ 
when the momentum-dependent symmetry potential is used? To answer
this question we have compared calculations using the MDI(1) and MDI(0) parameter sets 
with the momentum-dependent and -independent symmetry potentials. 
Using the potential in eq.\ \ref{mdi} these observables are actually 
more sensitive to the variation of $E_{sym}(\rho)$ than the calculations with the 
momentum-independent symmetry potentials. For instance, the slope of the differential 
flow $F^x_{n-p}(y)$ at y=0 changes from -1.9 to 21.4 MeV/c using eq. \ref{mdi} by changing 
from the MDI(1) to the MDI(0) parameter set, while it changes from 26 to 39 MeV/c with the 
momentum-independent symmetry potentials. The net change due to the variation of the $E_{sym}(\rho)$ 
is thus larger with the momentum-dependent symmetry potentials. 
  
In conclusion, the momentum-dependence of the symmetry potential is found to play an important
role in heavy-ion collisions induced by neutron-rich nuclei at RIA energies.
Symmetry potentials with and without the momentum-dependence but corresponding to the same
symmetry energy $E_{sym}(\rho)$ lead to significantly different predictions 
on several experimental observables that were previously identified as promising probes of the
$E_{sym}(\rho)$. With the momentum-dependent symmetry potential 
these observables are more sensitive to the change of $E_{sym}(\rho)$.  
The momentum-dependence of the symmetry potential is thus critical for investigating accurately 
the ${\rm EOS}$ of dense neutron-rich matter at RIA energies. 

We thank I. Bombaci, P. Danielewicz and W.G. Lynch for very helpful discussions. 
We would also like to thank C.M. Ko for a critical reading 
of the manuscript. This research is supported in part by the National Science Foundation of the 
United States under grant No. PHY-0088934 and PHY-0243571, the Natural Sciences and 
Engineering Research Council of Canada, and the Fonds Nature et Technologies of Quebec.

\end{document}